\documentclass[conference]{IEEEtran}

\ifCLASSOPTIONcompsoc
  \usepackage[nocompress]{cite}
\else
  \usepackage{cite}
\fi

\ifCLASSINFOpdf
\else
\fi

\usepackage{graphicx}
\usepackage[frozencache=true,cachedir=.]{minted}

\makeatletter
\let\@float@c@listing\@caption
\makeatother

\begin{document}
\title{Exposing the hidden layers and interplay in the quantum software stack}

\author{
\IEEEauthorblockN{Vlad Stirbu}
\IEEEauthorblockA{University of Jyväskylä\\
Jyväskylä, Finland\\
vlad.a.stirbu@jyu.fi}
\and
\IEEEauthorblockN{Arianne Meijer--van de Griend}
\IEEEauthorblockA{University of Helsinki\\
Helsinki, Finland\\
ariannemeijer@gmail.com}
\and
\IEEEauthorblockN{Jake Muff}
\IEEEauthorblockA{VTT Technical Research Centre of Finland\\
Espoo, Finland\\
jake.muff@vtt.fi}
}

\maketitle

\begin{abstract}
Current and near-future quantum computers face resource limitations due to noise and low qubit counts. Despite this, effective quantum advantage can still be achieved due to the exponential nature of bit-to-qubit conversion. However, optimizing the software architecture of these systems is essential to utilize available resources efficiently. Unfortunately, the focus on user-friendly quantum computers has obscured critical steps in the software stack, leading to ripple effects into the stack's upper layer induced by limitations in current qubit implementations. This paper unveils the hidden interplay among layers of the quantum software stack.
\end{abstract}

\IEEEpeerreviewmaketitle

\section{Introduction}

Quantum computers have demonstrated the potential to revolutionize various fields, including cryptography, drug discovery, material science, and machine learning, by leveraging the principles of quantum mechanics. However, the current generation of quantum computers, known as noisy intermediate-scale quantum (NISQ) computers~\cite{preskill}, suffer from noise and errors, making them challenging to operate. 
Additionally, the development of quantum algorithms requires specialized knowledge not readily available to the majority of software professionals. %

Thus, quantum hardware manufacturers tend to focus their software to be easy to use without specialized knowledge. On the one hand, this allows software engineers to easily use quantum computers because the intricacies of the hardware are abstracted away. On the other hand, the abstraction obscures many relevant layers in the quantum stack, such as circuit optimization, error mitigation, qubit routing, and pulse-level optimization, to name a few. Because they are obscured, the user has no good visibility what is executed on the quantum device, nor how this is done.

These abstraction layers are present in the classical computing as well, but there we have robust translation layers and ample computing resources to spare. Poorly optimised code is tolerable, because the compiler fixes many mistakes and the remaining inefficiencies can be handled at the expense of compute power. Only in specialized tasks do we need efficient code, and even then, we rarely write in Assembly. On the contrary, quantum computers are at a stage where we do not have these luxuries in quantum software, with quantum computing resources sparse and compilation tools and abstraction layers in their infancy.

Consequently, to effectively utilize the available hardware, it is crucial to provide users and algorithm designers with information about the hardware they are using such as the calibration state of the system, when it was last calibrated, the operations that can be performed on the device, and the topology of the qubits. This information needs to be presented to the user in such a way that it can be readily used when designing a quantum program. Revealing this information necessitates careful consideration for the introduction of new interfaces that maintain the layered structure of the quantum computer's software architecture~\cite{Foot00a}.

In this paper, we highlight the areas that need to be improved to make effective use of the capabilities of quantum computers and an overview of how to tackle these problems.

\section{The hidden interplay of layers}
Currently, any software that utilizes quantum hardware contains some part that is executed on a classical computer. In the most basic case, this classical part is the API instructing the quantum computer to execute a quantum algorithm, but the quantum algorithm can also be a component in a larger classical system~\cite{Montanaro2016}. Such an API is often part of a quantum development toolkit (QDK), for example Cirq\footnote{https://quantumai.google/cirq}, Qiskit\footnote{https://qiskit.org} or PennyLane\footnote{https://pennylane.ai}.

As an example case study, Listing \ref{listing:program} outlines a simplified program designed to run on VTT Q5 (Helmi)\footnote{https://vttresearch.github.io/quantum-computer-documentation/helmi/}. %
Utilizing the Qiskit framework, the program comprises four main components: the selection of the \textit{hard-coded, static backend}, describing the target hardware (lines 4-5), the \textit{assembly-style quantum circuit}~\cite{openqasm} (lines 7-10), the transpilation of the circuit into the machine-specific circuit (line 12), and the execution of the circuit on the provided backend (line 14).

\begin{listing}[t]
\begin{minted}[fontsize=\scriptsize,linenos=true,highlightlines={}]{python}
from qiskit import QuantumCircuit, execute, transpile
from iqm.qiskit_iqm import IQMProvider
# hardcoded backend
provider = IQMProvider("https://qc.vtt.fi/helmi/cocos")
backend = provider.get_backend()
# assembly style
circuit = QuantumCircuit(2, 2, name='Bell pair circuit')
circuit.h(0)
circuit.cx(0, 1)
circuit.measure_all()
# transpiling needs obscured
optimised_circuit = transpile(circuit, backend=backend)
# compiler and control hardware invisible
job = execute(optimised_circuit, backend, shots=1000)
result = job.result()
print(result)
\end{minted}
\caption{Simplified quantum program in Python}
\label{listing:program}
\end{listing}

\begin{figure}
    \centering
    \includegraphics[width=0.85\columnwidth]{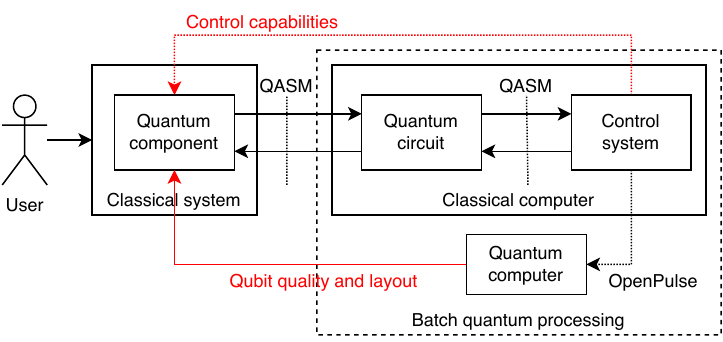}
    \caption{Quantum computing model: components and interfaces}
    \label{fig:enter-label}
\end{figure}

Despite the execution time on the hardware being extremely fast ($\mu s$), running a quantum program takes a long time.
A quantum program is \textit{executed multiple times} (i.e. shots) to judge the output distribution due to the probabilistic nature of noisy quantum computers~\cite{dalzell2023quantum}. This resembles \textit{batch processing}~\cite{batch-model} as depicted in Fig. \ref{fig:enter-label}. %

\textbf{Transpiling considerations.}
The task of the transpiler is to rewrite the quantum circuit to only contain instructions supported by the backend. This involves mapping the logical qubits to the on-device registers and dealing with the limited qubit connectivity of the backend (i.e. the qubit routing problem~\cite{qubit-routing-problem}).
Moreover, many quantum programs do not require every available qubit on a device~\cite{ichikawa2023comprehensive}, so the best qubit registers to use need to be selected by the transpiler.
Additionally, due to the noisy nature of NISQ hardware and the large variability of said noise~\cite{Smith:2022htg}, the transpiler needs to employ error mitigation strategies~\cite{Takagi_2022}. 
Additionally, which qubits on a particular quantum processing unit (QPU) are performing best also changes over time, with systems having to be re-calibrated frequently\cite{Smith:2022htg} so transpiling cannot be done in advance. 

\textbf{Transpiling is not compiling.}
Up until now, QDKs have incorrectly used transpiling as a synonym for compiling. A transpiler generates instructions in the same language as its input where a compiler translates it. Thus, \textit{the compiler is obscured}. In our example program, the transpiler generates another quantum circuit, but a compiler would generate quantum machine-code in the form of a schedule of control pulses to be sent to the hardware~\cite{openpulse}.
The side effect is that the \textit{compiler is naive}; it assumes that the transpiler found an optimal qubit mapping and replaces the instructions with equivalent pulses. However, the standard gates in the circuit model do not map one-to-one with pulses, leading to a prematurely optimized transpiler process that leaves performance on the table~\cite{earnest}.

\textbf{Control hardware visibility.}
A crucial part of the quantum stack that has not yet been mentioned in this description at all is the control hardware. 
A quantum computer cannot control itself, it is controlled by classical hardware that generates the pulses; all communication goes through the control hardware and accompanying control software. Additionally, the control hardware varies across vendors %
adding additional complexity, with the instruction set of the quantum computer fully characterized by the capabilities of the control hardware used to control the qubits. 
Yet, this last step in executing quantum software, \textit{the control hardware, is completely invisible} in the quantum stack. Particularly when there is much variability in its implementation, we cannot assume that all control hardware works the same way. Instead, we need to know the operating principles that govern the control hardware.

\section{Conclusion}
Although the stack's structure is largely appropriated from the classical stack, quantum computers are not classical computers, so these layers might have been defined prematurely.
To ensure the quantum stack is robust, we need to verify the assumptions on responsibilities, increase the transparency between layers, and enable communication between quantum hardware and quantum software via well-defined interfaces.

\bibliographystyle{ieeetr}
\bibliography{bibliography}

\end{document}